\documentclass[twocolumn]{aastex702}

\usepackage{lineno}
\linenumbers
\usepackage{rotating}

\usepackage{makecell}
\usepackage{comment}
\usepackage{natbib}
\usepackage{multirow}
\usepackage{booktabs}
\usepackage{epsfig}
\usepackage{rotating}
\usepackage{graphicx}
\usepackage{url}
\usepackage{xcolor}
\usepackage{amsmath}
\usepackage{color}
\usepackage{savesym}
\savesymbol{tablenum}
\usepackage{siunitx}
\usepackage[normalem]{ulem}
\usepackage{appendix}
\usepackage{enumerate}
\usepackage{subcaption}
\usepackage{gensymb}
\usepackage{xspace}
\usepackage{longtable}
\usepackage{afterpage}
\usepackage{threeparttable}
\usepackage{tabularx}
\usepackage[bottom]{footmisc}

\begin{document}

\title{The $M_{\rm BH}$--$R_{\rm b}$ relation and the high-mass end of the $M_{\rm BH}$--$\sigma$ relation }

\author[orcid=0000-0002-4140-0110,sname='Dullo']{Bililign T. Dullo}
\affiliation{Embry-Riddle Aeronautical University, Daytona Beach, FL 32114, USA}
\email[show]{dullob@erau.edu}

\begin{abstract}

  Using a sample of 151 galaxies with dynamically measured black hole
  (BH) masses ($M_{\rm BH}$), we investigate the scaling relations
  between $M_{\rm BH}$ and the stellar velocity dispersion, $\sigma$,
  and, for a subsample of 30 core-S\'ersic galaxies, between
  $M_{\rm BH}$ and the size of the partially depleted core,
  $R_{\rm b}$. Core-S\'ersic galaxies, identified using
  high-resolution \emph{Hubble Space Telescope} imaging and spanning
  both the normal-core ($R_{\rm b}<0.5$ kpc) and large-core
  ($R_{\rm b}>0.5$ kpc) regimes, define an updated
  \mbox{$M_{\rm BH}$--$R_{\rm b}$} relation of the form
  $M_{\rm BH} \propto R_{\rm b}^{1.16 \pm 0.10}$, with an rms scatter
  of $\Delta_{\rm rms} \simeq 0.28$ dex in $\log M_{\rm BH}$.  We
    find that S\'ersic and normal-core galaxies together follow a
    common log-linear $M_{\rm BH}$--$\sigma$ relation with a slope of
    $4.95 \pm 0.29$ and a scatter $\Delta_{\rm rms} \simeq 0.46$
    dex. Deviations from this relation arise at the highest BH
  masses, where large-core galaxies, including six with direct
  $M_{\rm BH}$ measurements, drive a significant upturn. We find that
  these galaxies typically host ultramassive black holes whose masses
  scale more strongly with $R_{\rm b}$ than $\sigma$, and lie
    $\sim ($1\text{--}4$) \times$ the intrinsic scatter (0.39 dex)
  above the relation defined by S\'ersic and normal-core galaxies. The
  $M_{\rm BH}$--$R_{\rm b}$ relation shows $\sim 30\text{--}47\%$ less
  scatter in $\log M_{\rm BH}$ than the corresponding
    $M_{\rm BH}$--$\sigma$ relation for the same sample. We interpret
  the high-mass upturn in the $M_{\rm BH}$--$\sigma$ diagram as a
  consequence of successive major, dry mergers, a scenario that
  naturally explains the observed flattening of the $\sigma-L_V$
  relation at $M_V \la -23.5$ mag.

   \end{abstract}


\section{Introduction}

Observations have shown that all massive elliptical galaxies and
massive bulges of disk galaxies host central black holes (BHs) with
masses \mbox{$M_{\rm BH} \sim 10^{6}-10^{10} M_{\sun}$}
\citep{1995ARA&A..33..581K,1998AJ....115.2285M,2005SSRv..116..523F}.
The masses of such black holes are known to correlate with several
host galaxy properties, including: stellar velocity dispersion,
$\sigma$
\citep{2000ApJ...539L...9F,2000ApJ...539L..13G,2013ApJ...764..184M},
spheroid luminosity, $L$ (e.g.,
\citealt{1995ARA&A..33..581K,2002MNRAS.331..795M}), dynamical mass
(e.g.,
\citealt{1998AJ....115.2285M,2003ApJ...589L..21M,2004ApJ...604L..89H})
and galaxy UV-[3.6] color \citep{2020ApJ...898...83D}.

In the standard cosmological paradigm, galaxies grow hierarchically
through the merger of smaller systems
\citep{1978MNRAS.183..341W,1991ApJ...379...52W,2001ApJ...561..517K}.
The ubiquity of central supermassive black holes (SMBHs)---combined
with the rarity of binary SMBHs---suggests that progenitor black holes
have coalesced in most merged galaxies. During major dry
(dissipationless) mergers, inspiralling SMBH binaries transfer orbital
angular momentum to surrounding stars, ejecting them from the central
regions and producing partially depleted stellar cores in luminous
($M_{V} \la -21.50 \pm 0.75$ mag) core-S\'ersic galaxies
\citep[e.g.,][]{1980Natur.287..307B,1991Natur.354..212E,2001ApJ...563...34M,
  2006ApJ...648..976M,2012ApJ...744...74G,2013ApJ...773..100K,2015ApJ...810...49V,2018ApJ...864..113R,2020MNRAS.497..739N}.
Consistent with this picture, careful modeling of core-S\'ersic
galaxies using HST imaging have revealed a strong correlation between
black hole mass and the size of depleted cores, as measured by the
break radius, $R_{\rm b}$
\citep[e.g.,][]{2007ApJ...662..808L,2012ApJ...755..163D,2014MNRAS.444.2700D,2019ApJ...886...80D,2021ApJ...908..134D}.


The tight $M_{\rm BH}$--$R_{\rm b}$ relation, which holds across the full
mass range of core-S\'ersic spheroids, has been largely overlooked,
despite displaying a lower level of scatter than the
$M_{\rm BH}$--$\sigma$ relation
\citep[][]{2019ApJ...886...80D,2021ApJ...908..134D}. The
$M_{\rm BH}-\sigma$ relation \citep[e.g.,][]{2016ApJ...831..134V},
which exhibits an upturn at the high mass end, predicts that SMBH
masses in the most massive galaxies (i.e.,
\mbox{$\sigma \sim 300 -390$ km s$^{-1}$},
\citealt{2003ApJ...594..225S,2007ApJ...662..808L,2007AJ....133.1741B,2014ApJ...797...82L})
cannot exceed $M_{\rm BH} \sim 6 \times 10^{9} M_{\sun}$. However,
ultramassive black holes (\mbox{UMBHs;
  $M_{\rm BH} \ga 10^{10} M_{\sun}$}) are increasingly discovered at
the centers of extremely massive present-day galaxies
\citep[e.g.,][]{2011Natur.480..215M,2016Natur.532..340T,2019ApJ...887..195M,2024ApJ...960..110S,2024MNRAS.530.1035D}.
\citet{2021ApJ...908..134D} found that the $M_{\rm BH}-\sigma$
relation underestimates the black hole mass for the most massive
galaxies by up to a factor of 40 (see also
\citealt{2019ApJ...886...80D}). Furthermore, predicted $M_{\rm BH}$
from the $M_{\rm BH}$--$R_{\rm b}$ and $M_{\rm BH}$--$L$ relations can
exceed $10^{10} M_{\sun}$
(\citealt{2007ApJ...662..808L,2019ApJ...886...80D,2021ApJ...908..134D}).

The upturn of the $M_{\rm BH}$--$\sigma$ relation at the highest $\sigma$
is consistent with substructure observed in the well-known
\citet[][$\sigma \propto L^{1/4}$]{1976ApJ...204..668F} relation for
core-S\'ersic galaxies.  \citet{2019ApJ...886...80D} separated
core-S\'ersic galaxies into normal-core ($R_{\rm b} < 0.5$ kpc) and
large-core ($R_{\rm b} \ga 0.5$ kpc) galaxies. They found that the
$\sigma$--$L_{V}$ relation flattens and exhibits larger scatter for
large-core spheroids with \mbox{$M_{V} \la -23.50 \pm 0.10$ mag},
$M_{*} \ga 10^{12}M_{\sun}$ and $R_{\rm e} \ga 10$ kpc (see also
\citealt{1991ApJ...375...15O,2006MNRAS.369.1081B,2007ApJ...662..808L,2012MNRAS.424..224H,2018MNRAS.474.1342M,2019ApJ...886...80D,2021ApJ...908..134D,2025ApJ...978...48S}). For
the combined sample of normal- and large-core galaxies, the slope of
the $\sigma$--$L$ relation, $1/(5.00 \pm 0.63)$, is shallower than that
of normal-core spheroids alone, $1/(3.50\pm 0.61)$; see also
\citealt{1981ApJ...251..508M,2007ApJ...662..808L,2013ApJ...769L...5K}).
At lower luminosities ($M_{V} \ga -21.5$ mag), coreless S\'ersic
galaxies follow a steeper relation, \mbox{$\sigma \propto L^{1/2}$}
(e.g., \citealt{1992AJ....103..851H,2005MNRAS.362..289M}).

\begin{figure*}
\hspace*{.50cm}  
\includegraphics[angle=0,scale=0.65]{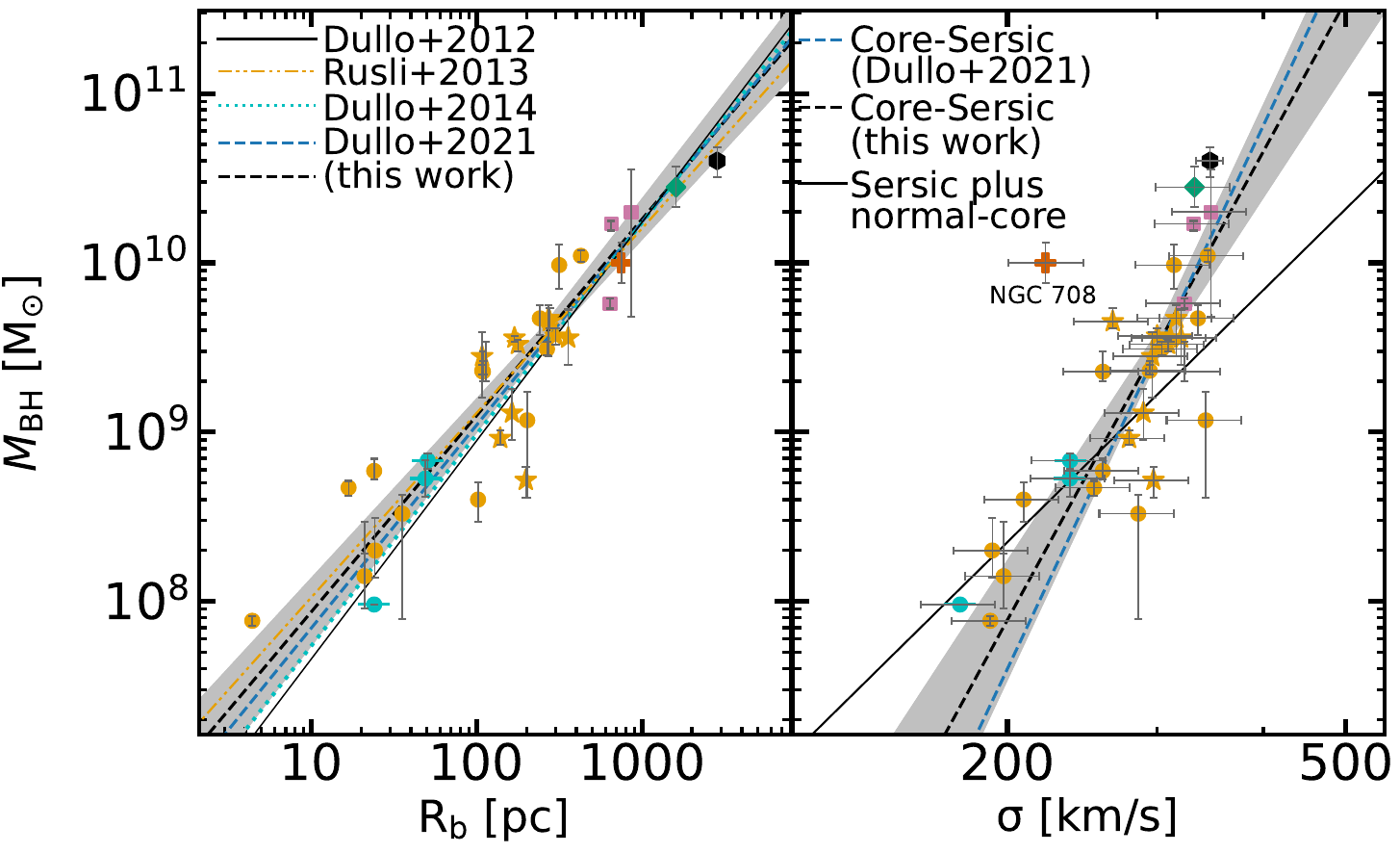}
\vspace*{.2172599cm}
\caption{ Relation between SMBH mass ($M_{\rm BH}$) and (a) the
    core-S\'ersic break radius ($R_{\rm b}$) and (b) central velocity
    dispersion ($\sigma$) for a sample of 30 core-S\'ersic galaxies
    with dynamically determined $M_{\rm BH}$. The black dashed lines
    are our symmetric {\sc bces} bisector regression fits
    (Table~\ref{Tab4}).  Filled orange circles and cyan disk symbol
    represent the 14 normal-core (i.e., $R_{\rm b} < 0.5$ kpc)
    elliptical galaxies and the three normal-core lenticular (S0)
    galaxies from \citet[their Table~2]{2014MNRAS.444.2700D},
    respectively, while filled purple boxes denote the three
    large-core (i.e., $R_{\rm b} > 0.5$ kpc) elliptical galaxies from
    \citet{2019ApJ...886...80D}. The 10 normal-core elliptical
    galaxies from \citet{2013AJ....146..160R} are indicated by filled
    orange stars.  The shaded regions show the 1$\sigma$ uncertainty
    for the regression fits (black dashed lines).  While we show the
    uncertainties on $R_{\rm b}$, they are smaller than the symbol
    sizes.  The dash-dot-dot line (left panel) represents the
    $M_{\rm BH}-R_{\rm b}$ relation from \citet{2013AJ....146..160R}.
    The black solid line (right panel) shows the symmetric {\sc bces}
    bisector fit to our S\'ersic + normal-core sample with dynamically
    determined $M_{\rm BH}$. Including 4C+37.11 (diamond,
    \citealt{2024ApJ...960..110S}), Holm 15A (hexagon,
    \citealt{2019ApJ...887..195M}), and NGC 708 (cross,
    \citealt{2024MNRAS.530.1035D})---excluded from our primary sample
    of 30 core-S\'ersic galaxies (see the text for details)---does not
    change the $M_{\rm BH}-R_{\rm b}$ relation.}  
\label{Fig1} 
\end{figure*}

 In a continued effort to investigate the $M_{\rm BH}$--$R_{\rm b}$
relation, and the offset at the high-mass end of the
$M_{\rm BH}$--$\sigma$ relation, we expand the sample of core-S\'ersic
galaxies with dynamically measured black hole masses 
presented in \citet{2021ApJ...908..134D} by $\sim$ 25\%.  Our full sample of 79
core-S\'ersic galaxies represents the largest to date used to study the
$M_{\rm BH}$--$R_{\rm b}$ and $M_{\rm BH}$--$\sigma$ relations for
core-S\'ersic galaxies.

\mbox{Section~\ref{Sec2}} discusses our sample of 151 galaxies with
dynamically measured SMBH masses, coupled with the 49 core-S\'ersic
galaxies with predicted $M_{\rm BH}$. \mbox{Section~\ref{Sec3}}
discusses the linear regression methods adopted.  In
\mbox{Section~\ref{Sec4.1}}, we compare the strengths of the
$M_{\rm BH}$--$R_{\rm b}$ and core-S\'ersic $M_{\rm BH}$--$\sigma$
relations.  Dividing our sample into S\'ersic, normal-core and
large-core galaxies, in Sections~\ref{Sec4.2} and \ref{Sec4.2b} we reveal substructure in
the $M_{\rm BH}$--$\sigma$ diagram.  In Section~\ref{Sec4.3}, we present
$M_{\rm BH}$ predicted using the \mbox{ $M_{\rm BH}$--$\sigma$} and
\mbox{$M_{\rm BH}$--$L$} relations and discuss underestimated BH masses of
large-core galaxies in the \mbox{$R_{\rm b}$--$M_{\rm BH}$} diagram. Section~\ref{Sec5} discusses the dry merger scenario that has been
presented in the literature to explain the upturn at the high-mass end of the
\mbox{$M_{\rm BH}$--$\sigma$} relation. In
\mbox{Section~\ref{Conc}}, we summarize our main conclusions. 
We include an appendix at the end of this paper presenting the BH 
scaling relations derived from our forward {\sc bces} and {\sc linmix} (Y$|$X) regression fits. 

\begin{table}
\caption{Core-S\'ersic galaxies with directly measured black hole masses.}
\label{Tab1}
\begin {minipage}{86mm}
\setlength{\tabcolsep}{0.0415863in}   
\begin{tabularx}{01.\textwidth}{@{}lllcccccc@{}}\hline
\hline
Galaxy&$D$&$\sigma$&$R_{\rm b}$ &log~($M_{\rm BH}/M_{\sun}$)\\
&(Mpc)&(km s$^{-1}$)&(kpc)&\\
(1)&(2)&(3)&(4)&(5)&\\
\multicolumn{1}{c}{} \\     
\toprule
IC~1459 &30.9 [1] &296&0.108 [1]	&	9.45$^{+0.14}_{-0.24}$ [1]   \\
NGC~0315 &67.0 [14]&304&0.266 [14]&9.49$^{+0.14}_{-0.24}$ [2]\\ 
NGC~0584  &19.6 [4]&198&0.021 [4]& 8.15$^{+0.13}_{-0.19}$  [3]  \\ 
NGC~1399   &19.4 [4]&342&0.202 [4] & 9.07$^{+0.7}_{-0.46}$  [4]  \\ 
NGC~1407 &28.1 [1]&266&0.276 [1]&	9.65$^{+0.08}_{-0.04}$ [1]  \\
NGC~1550 &51.6 [1]&300&0.300 [1] & 9.57$^{+0.05}_{-0.05}$ [1]  \\	
NGC~1600&66.0 [5]&331&0.650  [5] &10.23$^{+0.04}_{-0.04}$ [5] \\
NGC~3091&51.3 [1]&311&0.169 [1]&	9.56$^{+0.01}_{-0.03}$ [1]  \\
NGC~3379  &10.3 [4]&209&0.102 [4]& 8.60$^{+0.10}_{-0.13}$ [4]  \\ 
NGC~3608  &22.3 [4]&192&0.024 [4]&8.30$^{+0.19}_{-0.16}$  [4]  \\ 
NGC~3640&26.3 [4]&191&0.005 [4]&7.89$^{+0.14}_{-0.24}$ [3]  \\ 
NGC~3665 &32.1 [14]&237&0.049 [14]&8.73$^{+0.14}_{-0.24}$ [6]\\ 
NGC~3706&45.2 [4]&259&0.024 [4]&8.77$^{+0.14}_{-0.24}$ [7]\\ 
NGC~3842  &91.0 [4]&314&0.315 [4]& 9.98$^{+0.12}_{-0.14}$ [4]\\ 
NGC~4261 &31.6 [1]&297&0.198 [1]& 	8.72$^{+0.08}_{-0.10}$ [1]	\\
NGC~4291&25.5 [4]&285&0.036  [4]	 & 8.52$^{+0.11}_{-0.62}$  [4] \\ 
NGC~4374  &18.5 [1]&278&0.139 [1] &8.96$^{+0.05}_{-0.04}$ [1]\\
NGC~4382  &17.9 [4]&176&0.024 [4]&7.98$^{+0.05}_{-0.04}$ [8] \\
NGC~4472  &15.8 [4]&294&0.108 [4] &9.36$^{+0.04}_{-0.02}$ [1] \\
NGC~4486 &23.0 [5]&323& 0.640 [5] &$9.76^{+0.03}_{-0.03}$ [5] \\
NGC~4552 	 &14.9 [4]&253&0.017 [4]&8.67$^{+0.04}_{-0.05}$ [4]  \\ 
NGC~4649	&16.4 [4]&335&0.241 [4]&9.67$^{+0.08}_{-0.10}$ [4] \\ 
NGC~4889&96.6 [5]&347&0.860 [5] &10.30$^{+0.25}_{-0.62}$ [5]  \\
NGC~5328&64.1 [1]&316&0.271 [1]&	9.67$^{+0.08}_{-0.23}$ [1]\\
NGC~5516&58.4 [1]&309&0.178 [1]&	9.52$^{+0.03}_{-0.04}$ [1]	\\
NGC~5419&59.9 [4]&344&0.416 [4]&10.04$^{+0.03}_{-0.04}$ [9]	\\
NGC~5813	 &31.3 [4]&237& 0.051 [4]& 8.83$^{+0.04}_{-0.05}$ [4]\\ 
NGC~6086	&133.0 [1]&320&0.357 [1]	 &	9.56$^{+0.17}_{-0.16}$ [1]  \\	
NGC~7619	 &51.5 [4]&317& 0.109  [4]& 9.36$^{+0.06}_{-0.12}$ [10] \\  
NGC~7768 &112.8 [1]&289&0.164	 [1]&		9.11$^{+0.14}_{-0.16}$  [1]	\\
\mbox{NGC~708}$^{+}$ &68.5 [11] &222 [11] &  0.750 [11] &10.00$^{+0.12}_{-0.12}$ [11]\\
\mbox{4C+37.11}$^{+}$ &246.9 [12] &332 [12] & 1.603 [12] &10.45$^{+0.12}_{-0.12}$ [12]\\
\mbox{Holm 15A}$^{+}$ &252.8 [13] &346 [13] &  2.840 [13] &10.60$^{+0.08}_{-0.10}$ [13]\\ 
\hline
\end{tabularx}
\end {minipage}
Note. Col.\ (1) galaxy name. Col.\ (2) distance ($D$). Col.\ (3)
central velocity dispersion ($\sigma$) are taken from HyperLeda
\citep{2003A&A...412...45P}. Col.\ (4) core-S\'ersic break radius
($R_{\rm b}$). Col.\ (5) SMBH mass adjusted to our distance. The
subscript `$+$' indicates galaxies that are not in our primary
core-S\'ersic galaxy sample (see the text for details). Sources:
[1]=\citet[][]{2013AJ....146..160R}; [2]=\citet{2025ApJ...989...98P};
[3]= \citet{2019A&A...625A..62T}; [4]=\citet[][and references
therein]{2014MNRAS.444.2700D}; [5]=\citet[][and references
therein]{2019ApJ...886...80D}; [6]=\citet{2017MNRAS.468.4663O};
[7]=\citet{2014ApJ...781..112G}; [8]= \citet{2011ApJ...741...38G};
[9]=\citet{2023ApJ...950...15N}; [10]=\citet{2013AJ....146...45R};
[11]=\citet{2024MNRAS.530.1035D}; [12]=\citet{2024ApJ...960..110S};
[13]=\citet{2019ApJ...887..195M}; [14]=\citet{2023AA...675A.105D}.
  \end{table}

\begin{table}
\caption{Core-S\'ersic galaxies with predicted black hole masses.}
\label{Tab2}
\begin {minipage}{86mm}
\setlength{\tabcolsep}{0.0582863in}   
\begin{tabularx}{01.\textwidth}{@{}lllcccccc@{}}\hline
\hline
Galaxy&$D$&$\sigma$&$R_{\rm b}$ &$M_{V}$\\
&(Mpc)&(km s$^{-1}$)&(kpc)&(mag)\\
(1)&(2)&(3)&(4)&(5)&\\
\multicolumn{1}{c}{} \\     
\toprule
N0410 &72.4& 	        300   & 	0.178   &	-20.72 $\pm$	0.13 [1]	 \\     
N0507 &63.7&	 		292   & 	0.102 	&	-22.56 $\pm$	0.27 [2]\\         
N0741 &72.3&			287   & 	0.267 &	-23.39 $\pm$	0.18 [2]	\\     
N0777 &68.8&			324   & 	0.208 	&	-20.68 $\pm$	0.22 [1] \\     
N1016 &88.1&			288   & 	0.204 &	-23.31 $\pm$	0.22 [2]	 \\     
N1167 &68.0&			217   & 	0.045 		&    	-19.72 $\pm$	0.27 [1] \\     
N2300 &25.7&			266   & 	0.070  	&  	-21.33 $\pm$	0.20 [2]	 \\     
N2832 &105.0&			327   & 	0.483 	&	-24.58 $\pm$	0.23 [1]\\     
N3348 &41.8&			236   & 	0.066 &	-21.31 $\pm$	0.24 [1]	 \\     
N3613 &31.7&			220   & 	0.038 	&	-21.39 $\pm$	0.34 [1]	\\     
N4073 &85.3&			268  & 	0.090 	&	-23.42 $\pm$	0.37 [2]\\     
N4278 &15.6&			237  & 	0.052 	&	-20.91 $\pm$	0.27 [2]\\     
N4365 &19.9&			250  & 	0.127 	&	-22.02 $\pm$	0.28 [2]\\     
N4406 &16.7&			231  & 	0.061	&	-22.31 $\pm$	0.12  [2]	\\     
N4472 &15.8&			282 & 	0.108 	&	-22.68 $\pm$	0.33 [2]\\     
N4589 &21.4&			219 & 	0.027 	&	-21.04 $\pm$	0.24 [2]	\\     
N4874 &106.4&			272  & 	1.630 	&	-21.85 $\pm$	0.21 [3]\\     
N4914 &70.8&			225  & 	0.035 	&	-20.93 $\pm$	0.25 [1]	\\     
N5061 &32.6&			188  & 	0.034 	&	-22.46 $\pm$	0.21 [2]	\\     
N5322 &27.0&			230 &  	0.054 		&	-21.77 $\pm$	0.24 [2]\\     
N5557 &46.4&			259   & 	0.051 		&	-22.39 $\pm$	0.32 [2]\\     
N5631 &29.3&			168  & 	0.006 	&	-20.85 $\pm$	0.19 [1]	\\     
N5982 &41.8&			239 & 	0.051 	&	-22.08 $\pm$	0.27 [2]	\\     
N6166 &130.4&			300 & 	2.110 	&	-24.62 $\pm$	0.25 [3]\\     
N6849 &80.5&			202  & 	0.069 	&	-22.51 $\pm$	0.27 [2]	\\     
N6876 &54.3&			233 & 	0.119	&	-23.51 $\pm$	0.19  [2]\\     
4C+74.13	 &925.3&	        239 & 	2.240 	&	-24.12 $\pm$	0.20 [3]\\     
A0119BCG &185.7&		283 & 	0.670	&	-24.51 $\pm$	0.21 [3]\\     
A0168BCG$^{\dagger}$ &187.8 [4]&		253& 	0.071	&	-22.34 $\pm$	0.19\\     
A0189BCG$^{\dagger}$ &116.7 [4]&		212 & 	0.108	&	-22.55 $\pm$	0.17\\     
A0295BCG$^{\dagger}$ &185.0 [4]&		258 & 	0.477	&	-22.98 $\pm$	0.23\\    
A0419BCG$^{\dagger}$&136.5 [4]&		189 & 	0.139	&	-22.38 $\pm$	0.22\\            
\hline
\end{tabularx}
\end {minipage}
Note. Col.\ (1) galaxy name. Col.\ (2) distance ($D$). Col.\ (3)
central velocity dispersion ($\sigma$) are from HyperLeda. For the
BCGs, when HyperLeda $\sigma$ values are not available, we use those
from \citet{2014ApJ...797...82L}. Col.\ (4) core-S\'ersic break radius
($R_{\rm b}$). Col.\ (5) $V$-band spheroid absolute magnitude
($M_{V}$).  The core-S\'ersic break radii and $V$-band absolute
spheroid magnitudes were measured from PSF-convolved core-S\'ersic
fits to the galaxies' high-resolution HST surface brightness profiles
\citep{2014MNRAS.444.2700D,2019ApJ...886...80D, 2023AA...675A.105D}.
A `$\dagger$' indicates galaxies modeled in this work using HST WFPC2
light profiles. Sources for $D$, $R_{\rm b}$ and $M_{V}$ are:
[1]=\citet{2023AA...675A.105D}; [2]=\citet[][]{2014MNRAS.444.2700D};
[3]=\citet[][]{2019ApJ...886...80D};
[4]=\citet[][]{2003AJ....125..478L}; [5]=\citet{2017MNRAS.471.2321D}.
\end{table}

\begin{table}
\ContinuedFloat
\caption{-- continued.}
\label{Table0}
\begin {minipage}{86mm}
\setlength{\tabcolsep}{0.05082863in}   
\begin{tabularx}{01.\textwidth}{@{}lllcccccc@{}}\hline
\hline
Galaxy&$D$&$\sigma$&$R_{\rm b}$ &$M_{V}$\\
&(Mpc)&(km s$^{-1}$)&(kpc)&(mag)\\
(1)&(2)&(3)&(4)&(5)&\\
\multicolumn{1}{c}{} \\     
\toprule   
A0496BCG$^{\dagger}$  &143.0 [4]&		290 & 	0.427	&	-23.78 $\pm$	0.17\\     
A0912BCG$^{\dagger}$ &154.0 [4]&		207 & 	0.156	&	-22.62 $\pm$	0.23\\     
A1228BCG$^{\dagger}$&133.6 [4]&		277 & 	0.187	&	-22.88 $\pm$	0.20\\     
A1836BCG$^{\dagger}$  &171.0 [4]&		331 & 	0.249	&	-23.61	 $\pm$	0.20	\\     
A1983BCG$^{\dagger}$&154.1 [4]&		304 & 	0.209	&	-22.95 $\pm$	 0.19\\     
A2029BCG &363.0&		386 & 	4.200	&	-23.80	 $\pm$ 0.23 [5]\\\     
A2147BCG &153.4&		280 & 	1.280	&	-23.61 $\pm$	0.19 [3]\\\     
A2261BCG &958.8&		387 & 	3.333	&	-24.54 $\pm$	0.21\\     
A3376BCG$^{\dagger}$ &208.0 [4]&		319 & 	1.551	&	-23.62 $\pm$	0.25\\     
A3395BCG$^{\dagger}$ &217.0  [4]&		287 & 	0.341	&	-24.42 $\pm$	0.19\\     
A3528BCG$^{\dagger}$ &246.0 [4]&		364 & 	0.641	&	-23.21	$\pm$ 0.18\\     
A3556BCG$^{\dagger}$ &218.0 [4]&		347 & 	0.383	&	-24.10 $\pm$	0.25\\     
A3558BCG &204.9&		282 & 	1.300	&	-25.40	  $\pm$  0.23 [3]\\\     
A3562BCG &213.3&		263 & 	0.640	&	-23.42 $\pm$	0.26 [3]\\\     
A3571BCG &169.0&		313 & 	1.330	&	-23.35 $\pm$	0.20 [3]\\\     
A3716BCG &246.0&		275 & 	0.497	&	-23.35 $\pm$	0.21 [3]\\\     
A4059BCG$^{\dagger}$ &214.0 [4]&		289 & 	0.853	&	-23.64 $\pm$	0.25 \\     
\hline
\end{tabularx}
\end {minipage}
\end{table}   

\begin{table}
\caption{Large-core galaxy data}
\label{Tab3}
\begin {minipage}{220mm}
\setlength{\tabcolsep}{0.042818in}   
\begin{tabularx}{0.39\textwidth}{@{}lllcccccccccccccccc@{}}\hline
\hline
&\multicolumn{8}{c}{Predicted black hole masses}&\\
&&log~($M_{\rm BH}/M_{\sun}$)&\\
Galaxy&($\sigma$-based)&(L-based)&($R_{\rm b}$-based)&\\
(1)&(2)&(3)&(4)\\         
\toprule
NGC~4874      & 8.91$^{+0.42}_{-0.42}$&9.02$^{+0.34}_{-0.34}$& 10.50$^{+0.45}_{-0.45}$\\ 
NGC~6166      &9.19$^{+0.43}_{-0.43}$&10.50$^{+0.47}_{-0.47}$& 10.60$^{+0.46}_{-0.46}$ \\
4C+74.13  &8.68$^{+0.41}_{-0.41}$&10.23$^{+0.41}_{-0.41}$&  10.65$^{+0.46}_{-0.46}$\\
A0119    &9.06$^{+0.43}_{-0.43}$&10.45$^{+0.46}_{-0.46}$&10.12$^{+0.42}_{-0.42}$ \\  
A2029    &9.74$^{+0.47}_{-0.47}$& 10.06$^{+0.40}_{-0.40}$  &10.97$^{+0.44}_{-0.44}$\\
A2147   &9.03$^{+0.42}_{-0.42}$& 9.95$^{+0.41}_{-0.41}$ &10.38$^{+0.44}_{-0.44}$\\
A2261  &9.75$^{+0.48}_{-0.48}$&10.46$^{+0.47}_{-0.47}$&10.80$^{+0.45}_{-0.45}$ \\ 
A3376&9.32$^{+0.41}_{-0.41}$&9.96$^{+0.47}_{-0.47}$&10.49$^{+0.43}_{-0.43}$\\
A3528 &9.61$^{+0.44}_{-0.44}$&9.74$^{+0.47}_{-0.47}$&10.09$^{+0.41}_{-0.41}$\\
A3558    &9.05$^{+0.41}_{-0.41}$&10.92$^{+0.54}_{-0.54}$&10.40$^{+0.44}_{-0.44}$\\
A3562   &8.90$^{+0.41}_{-0.41}$&9.85$^{+0.39}_{-0.39}$&10.09$^{+0.42}_{-0.42}$\\ 
A3571   &9.28$^{+0.44}_{-0.44}$&9.81$^{+0.39}_{-0.39}$&10.41$^{+0.43}_{-0.43}$\\
A3716&8.99$^{+0.42}_{-0.42}$&9.81$^{+0.54}_{-0.54}$&9.91$^{+0.42}_{-0.42}$\\
A4059&9.10$^{+0.42}_{-0.42}$&9.97$^{+0.54}_{-0.54}$&10.19$^{+0.43}_{-0.43}$\\
\hline
&\multicolumn{8}{c}{Dynamically determined black hole masses}&\\
Galaxy&log~($M_{\rm BH}/M_{\sun}$)&Galaxy&log~($M_{\rm BH}/M_{\sun}$)\\
\hline
NGC 708 &10.00$^{+0.12}_{-0.12}$&NGC  4889& 10.30$^{+0.25}_{-0.62}$\\
NGC 1600&10.23$^{+0.04}_{-0.04}$&4C+37.11&10.45$^{+0.12}_{-0.12}$\\
NGC  4486& $9.76^{+0.03}_{-0.03}$&Holm 15A&10.60$^{+0.08}_{-0.10}$\\
\hline
\hline
\end{tabularx}
\end {minipage}
Note. Col.\ (1) galaxy name. Cols.\ (2--4) SMBH and UMBH masses were
predicted using the $M_{\rm BH}$--$\sigma$, $M_{\rm
  BH}$--$L_{\rm sph}$, and $M_{\rm BH}$--$R_{\rm b}$ relations,
respectively, for our sample of 14 large-core galaxies without
dynamically determined SMBH mass measurements (Section~\ref{Sec2.1}).
The lower part of the table lists six galaxies\tablenotemark{a} with dynamically
determined black hole masses reported to be large-core galaxies in the
literature. (see the text for details).  

\tablenotetext{a}{We note that three of these six galaxies with directly
measured BH masses (NGC 1600, NGC 4486, and NGC 4889) are already
included in Table~\ref{Tab1}.}

\end{table}

\begin{figure}
\hspace*{-.60cm}  
\includegraphics[angle=0,scale=0.461]{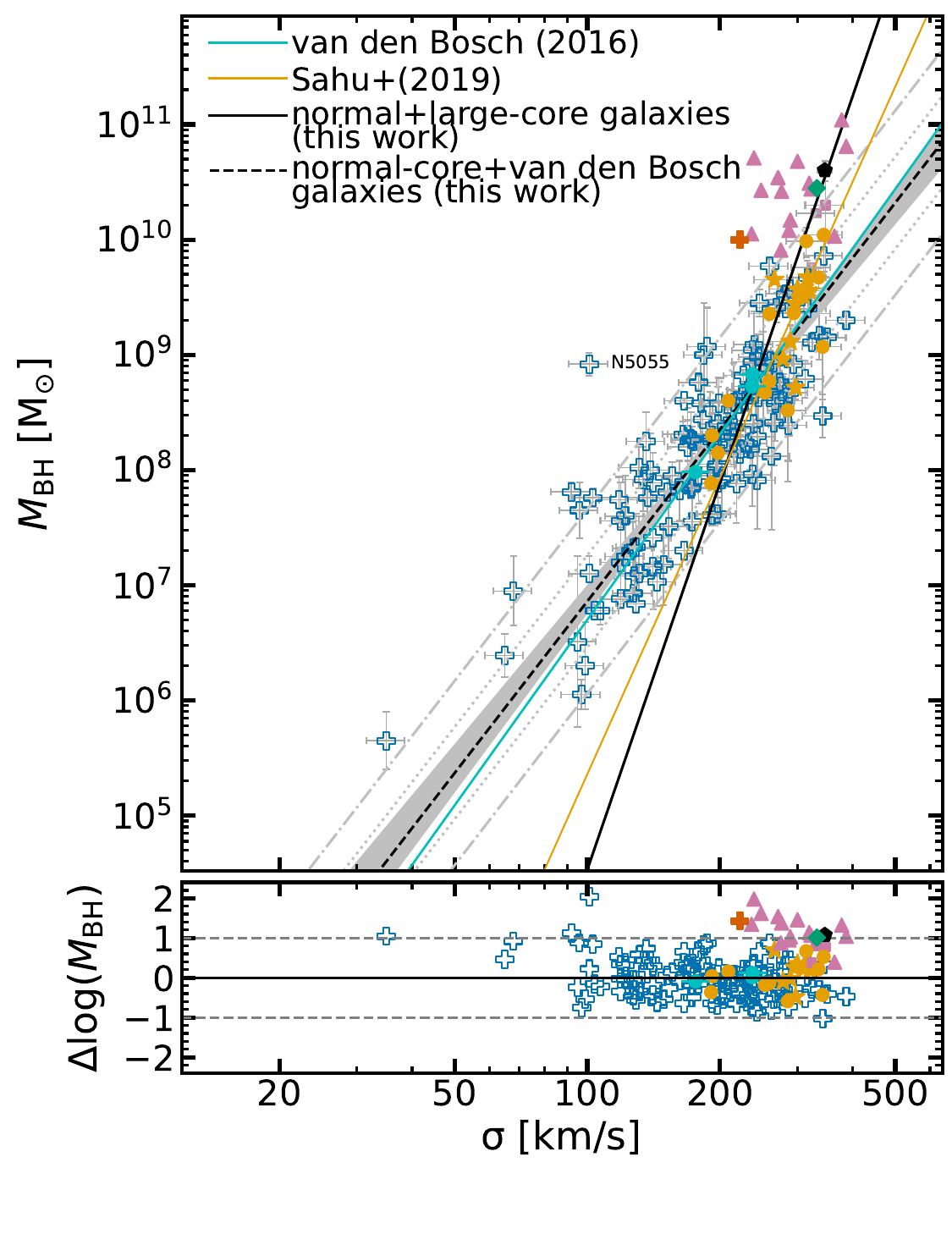}
\vspace*{-.872599cm}
\caption{ SMBH mass ($M_{\rm BH}$) plotted as a function of
    central velocity dispersion ($\sigma$). Similar to
    Fig.~\ref{Fig1}(b) but here we also show the 121 galaxies with
    dynamically determined SMBH masses from
    \citet{2016ApJ...831..134V} that are not in common with
    \citet{2013AJ....146..160R,2014MNRAS.444.2700D,2019ApJ...886...80D},
    open crosses.  In addition, we show 14 large-core galaxies with
    SMBH masses predicted using the \mbox{$M_{\rm BH}-R_{\rm b}$}
    relation (filled purple triangles; Table~\ref{Tab2}).  The black
    dashed line represents our symmetric {\sc bces} bisector fit to
    the ($M_{\rm BH}, \sigma$) data for the composite sample of 148
    non-(large-core) galaxies with directly measured BH masses---121
    galaxies from \citet{2016ApJ...831..134V}, 17 normal-core galaxies
    \citep{2014MNRAS.444.2700D} and 10 normal-core ellipticals
    \citep{2013AJ....146..160R}. The shaded region shows the 1$\sigma$
    uncertainty for the fit. The dotted and dashed-dotted lines show
    one and two times the measured vertical rms scatter in the
    \mbox{log $M_{\rm BH}$} direction ($\Delta_{\rm rms} = 0.46$ dex),
    respectively. The lower panel shows the residual profile about the
    fit. The solid black line is the {\sc bces} bisector fit to the 44
    core-S\'ersic galaxies (i.e., 30 with directly measured
    $M_{\rm BH}$ plus 14 large-cores with $M_{\rm BH}$ predicted using
    the \mbox{$M_{\rm BH}-R_{\rm b}$} relation). The solid cyan line
    indicates the $M_{\rm BH}-\sigma$ relation found by
    \citet{2016ApJ...831..134V} for their full sample of 230 galaxies,
    while the solid orange line is the core-S\'ersic
    $M_{\rm BH}-\sigma$ relation reported by
    \citet{2019ApJ...887...10S}. }
\label{Fig2} 
\end{figure}

  \section{Sample and Data} \label{Sec2}
  
  We note that most of the data used in this work to study the BH
  scaling relations are published elsewhere
  \citep{2014MNRAS.444.2700D,2019ApJ...886...80D,2021ApJ...908..134D,2023AA...675A.105D}. 
  
  To identify the partially depleted cores for the large-core and
  normal-core galaxies, we fitted the core-S\'ersic model to the {\it
    HST} stellar light distributions of the spheroidal components of
  the galaxies
  (\citealt{2014MNRAS.444.2700D,2015ApJ...798...55D,2019ApJ...886...80D,2021ApJ...908..134D,2023AA...675A.105D,2023MNRAS.522.3412D,2024MNRAS.532.4729D}).

  Using a sample of 24 core-S\'ersic galaxies with direct ly measured
  SMBH masses and 27 core-S\'ersic with predicted SMBH masses,
  \citet{2021ApJ...908..134D} investigated the $M_{\rm BH}-R_{\rm b}$
  relation, and the offset at the high-mass end of the
  $M_{\rm BH}-\sigma$ relation.  Here, we expand the sample of
  core-S\'ersic galaxies with direct SMBH masses from 24 to 30,
  representing a 25\% increase, and increase the number of
  core-S\'ersic galaxies with predicted SMBH masses from 27 to 49, an
  $\sim$82\% increase (Tables~\ref{Tab1} and \ref{Tab2}).  Our full
  sample of 79 core-S\'ersic galaxies are compiled by combining the
  data from \citet{2014MNRAS.444.2700D}, \citet{2019ApJ...886...80D}
  \citet{2023AA...675A.105D} and 13 core-S\'ersic galaxies in this
  work (Table~\ref{Tab2}).  Of the 30 (49) core-S\'ersic galaxies with
  directly measured (predicted) SMBH masses, three (14) have large
  cores (i.e., $R_{\rm b} > 0.5$ kpc).
  
  We also present the $M_{\rm BH}-R_{\rm b}$ and $M_{\rm BH}-\sigma$
  relations, including three galaxies recently reported to host large
  cores and ultramassive black holes: 4C+37.11
  \citep{2024ApJ...960..110S}, Holm 15A \citep{2019ApJ...887..195M},
  and NGC 708 \citep{2024MNRAS.530.1035D}. Core-S\'ersic galaxies
  possess a central stellar light (or mass) deficit relative to the
  inward extrapolation of their outer
  \citet{1963BAAA....6...41S,1968adga.book.....S} $R^{1/n}$ profile,
  resulting in shallow inner slopes ($\gamma \lesssim 0.3$). However,
  Holm 15A and NGC 708 were not identified as having such deficits
  \citep{2019ApJ...887..195M,2024MNRAS.530.1035D}, while 4C+37.11 was
  modeled with a core-S\'ersic model with a sharper transition
  ($\alpha = 100$) and a relatively steep inner slope,
  $\gamma \sim 0.44$, \citep{2024ApJ...960..110S}. 
  
For Holm 15A,  \citet{2019ApJ...887..195M} reported  the so-called `cusp radius', 
defined as the radius where the local logarithmic slope $\gamma = 1/2$ \citep{1997ApJ...481..710C}. 
We note that even galaxies without depleted cores can have a radius at which 
$\gamma = 1/2$. For NGC 708, \cite{2024MNRAS.530.1035D} fitted the inner part 
of the spheroid using a core-S\'ersic  model with a low S\'ersic index ($n \sim 1.8$) 
and a small effective radius $R_{\rm e} \sim 0.5$ kpc.  However, the break radius is known 
to be sensitive to the adopted fitted range, particularly when the light profile does 
not cover the full radial extent of the spheroid \citep{2012ApJ...755..163D}. 
  
  Nevertheless, using HyperLeda photometric and kinematic
  measurements, we find that all three galaxies fulfill the criteria commonly associated with
  core-S\'ersic galaxies, namely high luminosities
  ($M_V \lesssim -21.5$ mag) and large stellar velocity dispersions
  ($\sigma \gtrsim 200~{\rm km~s^{-1}}$; see Dullo 2014; Dullo et
  al. 2023). We therefore include these galaxies in the comparison
  plots, while excluding them from our primary core-S\'ersic sample and from
  the baseline regression fits involving $R_{\rm b}$.

  In order to construct the $M-\sigma$ relations, we used a sample of
  121 dynamically determined SMBH masses from \citet[see their
  Table~2]{2016ApJ...831..134V}, excluding those in common with
  \citet{2014MNRAS.444.2700D}, \citet{2019ApJ...886...80D},
  \citet{2021ApJ...908..134D} and \citet{2023AA...675A.105D}.  The
  majority ($\sim 94$\%) of these galaxies are S\'ersic galaxies,
  while a small fraction ($\sim 6$\%), with
  $M_{\rm BH }\ga 2 \times 10^{9} M_{\sun}$ and
  \mbox{$\sigma \ga 280 $ km s$^{-1}$}, may be core-S\'ersic galaxies
  with $R_{\rm b} < 0.5$ kpc.

  Central stellar velocity dispersions ($\sigma$) for the sample
  galaxies were primarily obtained from
  HyperLeda\footnote{http://atlas.obs-hp.fr/hyperleda/}
  \citep{2003A&A...412...45P}. For the BCGs, when HyperLeda $\sigma$
  values are not available, we adopt those from
  \citet{2014ApJ...797...82L}. We assume a 10\% uncertainty on
  $\sigma$ when deriving the black hole scaling relations
  (Section~\ref{Sec4}).

\subsection{Predicted black hole masses for core-S\'ersic galaxies} \label{Sec2.1}

To better investigate the black hole scaling relations, we use the
velocity dispersion ($\sigma$) and our $V$-band spheroid absolute
magnitude ($M_{V}$) for 49 (35 normal-core plus 14 large-core)
galaxies (see Table~\ref{Tab2}) to predict their black hole masses. We
made use of the \citet[their Table 3]{2013ApJ...764..151G} non-barred
\mbox{$M_{\rm BH}-\sigma$} relation to predict the
\mbox{$\sigma$--based} SMBH masses. {The \mbox{$L_{\rm}$--based} SMBH
masses were predicted using the near-linear \citet[their Table
3]{2013ApJ...764..151G} $B$-band core-S\'ersic $M_{\rm BH}-L$ relation
transformed here into the $V$-band using total dust corrected $B-V$
galaxy color from HyperLeda. } We also use our
\mbox{$M_{\rm BH}-R_{\rm b}$} relation to predict black hole masses
for the 14 large-core galaxies in our sample with no directly measured
black hole masses (Table~\ref{Tab3}).

\begin{sidewaystable*}
\centering
\scriptsize
\caption{Black Hole Scaling Relations}
\label{Tab4}
\begin{tabular}{ccccccc}
\hline
Relation & BCES fit & $\Delta_{\rm rms}$ & $\epsilon$ & $r_s/P$ & $r_p/P$ & Sample \\
\hline
&&{\bf Dynamically measured SMBH masses only}&&&&\\
  &&{\bf Core-S\'ersic galaxies }&&&&\\
 $M_{\rm BH}-R_{\rm b}$
 &$\mbox{log}\left(\frac{M_{\rm BH}}{M_{\sun}}\right)= (1.16\pm
           0.10) \mbox{log}\left(\frac{R_{\rm b}}{\mbox{250 pc}}\right)$ +~($9.56 ~ \pm  0.05$)&0.28&0.27 $\pm$  0.05
 &  0.91/$5.6 \times 10^{-12}$&0.90/$1.1\times10^{-11}$&30 [a]\\
$M_{\rm BH}-\sigma $
        &$\mbox{log}\left(\frac{M_{\rm BH}}{M_{\sun}}\right)= (9.20\pm
            1.53) \mbox{log}\left(\frac{\sigma}{\mbox{300 {\rm km s$^{-1}$}}}\right)$ +~($9.51 ~ \pm  0.07$)& 0.40&0.20 $\pm$ 0.09
 &0.85/$1.8\times10^{-9}$&0.86/$9.9\times10^{-10}$&30 [a]\\
   &&{\bf Including 4C+37.11, Holm 15A, and NGC 708 }&&&&\\
 $M_{\rm BH}-R_{\rm b}$
 &$\mbox{log}\left(\frac{M_{\rm BH}}{M_{\sun}}\right)= (1.15\pm
           0.08) \mbox{log}\left(\frac{R_{\rm b}}{\mbox{250 pc}}\right)$ +~($9.55 ~ \pm  0.05$)&0.27&0.27 $\pm$  0.05
 &  0.93/$2.0 \times 10^{-14}$&0.92/$5.1\times10^{-14}$&33 [b]\\
$M_{\rm BH}-\sigma $
        &$\mbox{log}\left(\frac{M_{\rm BH}}{M_{\sun}}\right)= (10.22 \pm
            1.69) \mbox{log}\left(\frac{\sigma}{\mbox{300 {\rm km s$^{-1}$}}}\right)$ +~($9.61 ~ \pm  0.09$)& 0.51&0.31 $\pm$ 0.09
 &0.79/$3.4\times10^{-8}$&0.79/$4.7\times10^{-8}$&33 [b]\\
 &&{\bf S\'ersic plus Core-S\'ersic galaxies }&&&\\
 $M_{\rm BH}-\sigma $
 &$\mbox{log}\left(\frac{M_{\rm BH}}{M_{\sun}}\right)= (4.95 \pm 0.29)
 \mbox{log}\left(\frac{\sigma}{\mbox{200 {\rm km s$^{-1}$}}}\right)$
 +~($8.35 ~\pm 0.04$)& 0.46&0.39 $\pm$0.03
 &0.85/$ 2.1\times10^{-43}$&0.85/$  1.2\times10^{-43}$&148 [c] \\
 $M_{\rm BH}-\sigma $
 &$\mbox{log}\left(\frac{M_{\rm BH}}{M_{\sun}}\right)= (5.10 \pm 0.29)
 \mbox{log}\left(\frac{\sigma}{\mbox{200 {\rm km
         s$^{-1}$}}}\right)$+~($8.37 ~\pm 0.04$)& 0.47 &0.39 $\pm$0.03
 &0.86/$1.1\times10^{-45}$&0.85/$4.0\times10^{-44}$&151 [d]\\
    &&{\bf Including 4C+37.11, Holm 15A, and NGC 708 }&&&&\\
 $M_{\rm BH}-\sigma $
 &$\mbox{log}\left(\frac{M_{\rm BH}}{M_{\sun}}\right)= (5.28 \pm 0.31)
 \mbox{log}\left(\frac{\sigma}{\mbox{200 {\rm km
         s$^{-1}$}}}\right)$+~($8.39 ~\pm 0.04$)& 0.50 &0.42 $\pm$0.04
 &0.86/$1.4\times10^{-46}$&0.85/$1.0\times10^{-44}$&154 [e]\\\\   \\
&&{\bf Including predicted SMBH masses}&&&&\\
  &&{\bf Core-S\'ersic galaxies }&&&&\\
$M_{\rm BH}-\sigma $
        &$\mbox{log}\left(\frac{M_{\rm BH}}{M_{\sun}}\right)= (11.21 \pm
             1.58) \mbox{log}\left(\frac{\sigma}{\mbox{300 {\rm km s$^{-1}$}}}\right)$ +~($9.86~ \pm  0.11$)& 0.72&0.34 $\pm$ 0.12
 &0.56/$1.6 \times10^{-4}$&0.66/$1.0\times10^{-6}$&44 [f]\\
   &&{\bf Including 4C+37.11, Holm 15A, and NGC 708 }&&&&\\
  $M_{\rm BH}-\sigma $
        &$\mbox{log}\left(\frac{M_{\rm BH}}{M_{\sun}}\right)= (11.00 \pm
             1.26) \mbox{log}\left(\frac{\sigma}{\mbox{300 {\rm km s$^{-1}$}}}\right)$ +~($9.90~ \pm  0.10$)& 0.72&0.41 $\pm$ 0.11
 &0.57/$7.8 \times10^{-5}$&0.65/$2.5\times10^{-6}$&47 [g]\\ 
  &&{\bf S\'ersic plus Core-S\'ersic galaxies }&&&\\
  $M_{\rm BH}-\sigma $
 &$\mbox{log}\left(\frac{M_{\rm BH}}{M_{\sun}}\right)= (5.87\pm 0.37)
 \mbox{log}\left(\frac{\sigma}{\mbox{200 {\rm km s$^{-1}$}}}\right)$
 +~($8.45 ~\pm 0.05$)& 0.58&0.44 $\pm$0.04
 &0.85/$ 6.1\times10^{-47}$&0.82/$ 7.1\times10^{-42}$&165 [h]\\
    &&{\bf Including 4C+37.11, Holm 15A, and NGC 708 }&&&&\\
    $M_{\rm BH}-\sigma $
 &$\mbox{log}\left(\frac{M_{\rm BH}}{M_{\sun}}\right)= (5.99 \pm 0.37)
 \mbox{log}\left(\frac{\sigma}{\mbox{200 {\rm km s$^{-1}$}}}\right)$
 +~($8.47 ~\pm 0.05$)& 0.60&0.46 $\pm$0.04
 &0.86/$ 4.1\times10^{-48}$&0.83/$ 1.1\times10^{-43}$&168 [i]\\
\hline
\end{tabular}
Note---The different columns represent: the BH
   scaling relations, rms scatter in the vertical log $M_{\rm BH}$
   direction ($\Delta_{\rm rms}$), intrinsic scatter from the Bayesian
   {\sc linmix} fits ($\epsilon$), and the Spearman and Pearson
   correlation coefficients ($r_{\rm s}$ and $r_{\rm p}$,
   respectively) and the associated probabilities. For scaling
   relations including predicted SMBH masses, the SMBH masses were
   predicted from the $M_{\rm BH}-R_{\rm b}$ relation, see
   Table~\ref{Tab3}. Definitions of samples ([a] -- [i]) are presented
   in Table~\ref{Tab5}.
\end{sidewaystable*}


\begin{table*}
\centering
\caption{Definition of Galaxy Samples Used in Table~\ref{Tab4}}
\label{Tab5}

\begin{tabular}{cp{0.83\textwidth}}
\hline\hline
Sample & Description \\
\hline

[a] &
30 core-S\'ersic galaxies with dynamically determined SMBH masses:
15 normal-core galaxies from \citet{2014MNRAS.444.2700D},
10 normal-core ellipticals from \citet{2013AJ....146..160R},
3 large-core ellipticals from \citet{2019ApJ...886...80D},
and 2 normal-core galaxies from \citet{2023AA...675A.105D}. \\

[b] &
33 core-S\'ersic galaxies: the 30 galaxies in sample [a] plus
4C+37.11 \citep{2024ApJ...960..110S},
Holm 15A \citep{2019ApJ...887..195M}, and
NGC 708 \citep{2024MNRAS.530.1035D}. \\

[c] &
148 non-large-core galaxies: 27 normal-core galaxies from
sample [a] plus 121 galaxies with dynamically measured SMBH masses
from \citet{2016ApJ...831..134V} not in common with
\citet{2013AJ....146..160R,2014MNRAS.444.2700D,
2019ApJ...886...80D,2023AA...675A.105D}. \\

[d] &
151 galaxies: the 148 galaxies in sample [c] plus
3 large-core ellipticals from
\citet{2019ApJ...886...80D}. \\

[e] &
154 galaxies: the 151 galaxies in sample [d] plus
4C+37.11, Holm 15A  and
NGC 708. \\

[f] &
44 core-S\'ersic galaxies: the 30 galaxies in sample [a] plus
14 large-core galaxies with SMBH masses predicted from the
$M_{\rm BH}-R_{\rm b}$ relation (Table~\ref{Tab3}). \\

[g] &
47 core-S\'ersic galaxies: the 44 galaxies in sample [f] plus
4C+37.11, Holm 15A, and NGC 708. \\

[h] &
165 galaxies: the 151 galaxies in sample [d] plus
14 large-core galaxies with predicted SMBH masses from
Table~\ref{Tab3}. \\

[i] &
168 galaxies: the 154 galaxies in sample [e] plus
14 large-core galaxies with predicted SMBH masses from
Table~\ref{Tab3}. \\

\hline
\end{tabular}
\end{table*}

\section{ Regression Analysis}\label{Sec3}

 To derive the black hole scaling relations (Table~\ref{Tab4}), we
  performed linear regressions using the Bivariate Correlated Errors
  and intrinsic Scatter ({\sc bces}) code
  (\citealt{1996ApJ...470..706A}) and the Bayesian linear regression
  routine ({\sc linmix\_err}, \citealt{2007ApJ...665.1489K}). The
  fitted {\sc bces} bisector regressions, based on the python module
  by \citet{2012Sci...338.1445N}, are symmetrical and bisect the
  forward ($Y|X$) and inverse ($X|Y$) lines.  Both the {\sc bces} and
  {\sc linmix\_err} methods allow us to account for errors in the
  variables under consideration (e.g.,
  \citealt{2020ApJ...898...83D,2021ApJ...908..134D}). We note that a
  forward regression ($Y|X$) tend to yield a shallower slope than an
  inverse regression, $X|Y$ \citep{1992ApJ...397...55F}.  To address
  the `theorist’s question', i.e., the intrinsic correlations between
  two quantities, symmetrical regressions are generally preferred when
  the `dependent' and `independent' variables are not clearly defined
  \citep{2006ApJ...637...96N}. The {\sc bces} bisector regressions are
  presented throughout this paper (Table~\ref{Tab4}). Nevertheless,
  our symmetric {\sc bces} relations are consistent within the errors
  with the forward ($Y|X$) regressions obtained with {\sc bces} and
  {\sc linmix} (see Table~\ref{Tab4} and Appendix~\ref{App2}).
Additionally, we re-analyzed the data using the symmetric orthogonal
{\sc bces} regressions and obtained consistent results.

We quantify the strength of the correlations using the Spearman and
Pearson correlation coefficients and the associated $p$-values, for
which the null hypothesis that the two data sets are uncorrelated is
true.  The intrinsic scatter from our {\sc linmix} regression fits to
the ($M_{\rm BH}, R_{\rm b}$) and ($M_{\rm BH}, \sigma$) data sets
\citep[e.g.,][]{2020ApJ...898...83D}, $\epsilon$, as well as the
 vertical root-mean-square (rms) scatter in the log $M_{\rm BH}$
directions ($\Delta_{\rm rms}$), are presented in Table~\ref{Tab4}.

\section{Results}\label{Sec4}

\subsection{$M_{\rm BH}$--$R_{\rm b}$ and $M_{\rm BH}$--$\sigma$ relations for core-S\'ersic galaxies }\label{Sec4.1}   

Fig.~\ref{Fig1} plots the $M_{\rm BH}-R_{\rm b}$ and
$M_{\rm BH}-\sigma$ relations for our sample of 30 core-S\'ersic
galaxies with directly measured SMBH masses.  The intrinsic scatter in
the $M_{\rm BH}-R_{\rm b}$ relation is $\epsilon \sim 0.27 \pm
0.05$. The correlation between $M_{\rm BH}$ and $R_{\rm b}$ has a
total rms scatter in log $M_{\rm BH}$ direction of $\Delta \sim 0.28$
dex, Spearman correlation coefficient ($r_{\rm s}$)/$p$-value of
\mbox{$0.91$/$5.6\times10^{-12}$} and Pearson correlation coefficient
($r_{\rm p})/$$p$-value of \mbox{$0.90$/$1.1\times10^{-11}$}. In
contrast, the core-S\'ersic $M_{\rm
  BH}-\sigma$ relation constructed using the same sample
(Fig.~\ref{Fig1}) has $\epsilon \sim 0.25 \pm 0.05$, $\Delta \sim
0.40$ dex, $r_{\rm s}/$p-value \mbox{$0.85$/$1.8\times10^{-9}$} and
$r_{\rm p}/$p-value \mbox{$0.86$/$9.9\times10^{-10}$}
(Table~\ref{Tab4}).

The slope of the updated {$M_{\rm BH}-R_{\rm b}$ relation ($1.16 \pm
  0.10$) is in good agreement with those reported in previous studies,
  including \citet[][their Eqs. 11 and 12 with an average slope of
  $1.29 \pm 0.33$]{2012ApJ...755..163D}, \citet[][$1.09 \pm
  0.20$]{2013AJ....146..160R}, \citet[][$1.25 \pm
  0.25$]{2014MNRAS.444.2700D} and \citet[][$1.20 \pm
  0.25$]{2021ApJ...908..134D}.

  Using a sample of 33 core-S\'ersic galaxies, i.e., including
    three galaxies recently reported to host UMBHs and large cores
    (4C+37.11, \citealt{2024ApJ...960..110S}, Holm 15A,
    \citealt{2019ApJ...887..195M}, and NGC 708,
    \citealt{2024MNRAS.530.1035D}), we obtain a consistent relation
    (\mbox{$M_{\rm BH} \propto R_{\rm b}^{1.15 \pm 0.08}$}); see
    Tables~\ref{Tab3} and \ref{Tab4}. This relation has a total rms
    scatter in log $M_{\rm BH}$ direction of $\Delta \sim 0.27$ dex, a
    Spearman correlation coefficient ($r_{\rm s}$)/$p$-value of
    \mbox{$0.93$/$2.0\times10^{-14}$} and a Pearson correlation
    coefficient ($r_{\rm
      p})/$$p$-value of \mbox{$0.92$/$5.1\times10^{-14}$}.  The
    corresponding core-S\'ersic $M_{\rm
      BH}-\sigma$ relation (Fig.~\ref{Fig1}) has $\epsilon \sim 0.31
    \pm 0.09$, $\Delta \sim 0.51$ dex, $r_{\rm
      s}/$p-value \mbox{$0.79$/$3.4\times10^{-8}$} and $r_{\rm
      p}/$p-value \mbox{$0.79$/$4.7\times10^{-8}$}
    (Table~\ref{Tab4}).

Our sample consists of a total of six galaxies which host UMBHs
$M_{\rm BH} \ga 10^{10} M_{\sun}$, representing $\sim
18$\% of all the core-S\'ersic galaxies with measured $M_{\rm
  BH}$ to date.  Fig.~\ref{Fig1} shows that the large-core galaxies
(purple boxes, black hexagon, green diamond and orange cross) follow
the log-linear $M_{\rm BH}-R_{\rm
  b}$ relation traced by the relatively less massive, normal-core
population (filled orange circles, cyan disk symbol and filled orange
stars), see Tables~\ref{Tab1}, \ref{Tab3} and \ref{Tab4}.   The
  large break radii of all the six large-core galaxies with measured
  $M_{\rm
    BH}$ (4C+37.11, NGC~708, NGC~1600, NGC~4486, NGC~4889 and Holm
  15A) are in good agreement with the galaxies' large black hole
  masses (Fig.~\ref{Fig1}).

  We compare the $M_{\rm BH}-R_{\rm b}$ and $M_{\rm BH}-\sigma$
  relations constructed using the same sample (see
  Table~\ref{Tab4}). We find that the tight $M_{\rm BH}-R_{\rm b}$
  relation exhibits $\sim$ $30-47$\% less scatter in log $M_{\rm BH}$
  than the core-S\'ersic $M_{\rm BH}-\sigma$ relation, albeit the two
  relations exhibit a similar level of intrinsic scatter
  ($\epsilon \sim 0.26 \pm 0.06$).

\subsection{The $M_{\rm BH}$--$\sigma$ relation}\label{Sec4.2}

In order to investigate the upturn tendency of the
$M_{\rm BH}$--$\sigma$ relation at high $\sigma$, we derive its slopes
by dividing our sample into S\'ersic, normal-core and large-core
galaxies.  Fig.~\ref{Fig2} shows the best-fitting {\sc bces} bisector
regression for the sample of 148 galaxies (dashed line): 121 galaxies
from \citet{2016ApJ...831..134V} (most which are S\'ersic galaxies)
and 27 normal-core galaxies (Table~\ref{Tab2}). The shaded region
represents the 1$\sigma$ uncertainty of the fit. We find that
\mbox{non-(large-core)} (i.e., S\'ersic and normal-core) galaxies
define a single log-linear $M_{\rm BH}$--$\sigma$ relation with a
slope of $4.95 \pm 0.29$, \mbox{$\epsilon \sim $0.39 dex} and
\mbox{$\Delta_{\rm rms} \sim 0.46$ dex} in the \mbox{log
  $M_{\rm BH}$}, consistent with previous studies
\citep{2013ApJ...764..151G,2015MNRAS.446.2330S,2016ApJ...831..134V,2018MNRAS.473.5237K,2020ApJ...898...83D,2021ApJ...908..134D}. When
the 27 normal-core galaxies are excluded from the regression analysis,
the {\sc bces} bisector $M_{\rm BH}$--$\sigma$ relation for the 121
galaxies remains similar, with a slope of $\sim 4.66 \pm 0.29$ and an
intercept of $\sim -2.39 \pm 0.67$.

While the core-S\'ersic $M_{\rm BH}$--$\sigma$ relation for our
  sample of 30 core-S\'ersic galaxies with directly measured SMBH
  masses (Fig.~\ref{Fig1}) has a significantly steeper slope of
  $9.20 \pm 1.53$ and an intercept of $\sim -13.28 \pm 3.79$, we did
  not find evidence for a marked division between S\'ersic and
  normal-core galaxies, albeit see \cite{2019ApJ...887...10S}.  We
  have compared the relative quality of a single unified
  $M_{\rm BH}$--$\sigma$ relation with separate S\'ersic and
  normal-core relations using the Akaike Information Criterion (AIC,
  \citealt{1974ITAC...19..716A}) and Bayesian Information Criterion
  (BIC, \citealt{10.1214/aos/1176344136}).

\begin{equation}
\mathrm{AIC}
=
2k - 2\ln(\mathcal{L}),
\end{equation}

\begin{equation}
\mathrm{BIC}
=
k\ln(n) - 2\ln(\mathcal{L}),
\end{equation}
and a variation of the AIC, AICc can be written as
\begin{equation}
\mathrm{AICc}
=
\mathrm{AIC}
+
\frac{2k(k+1)}{n-k-1}
\end{equation}
where $\mathcal{L}$ is the maximum value of the likelihood function for the model, $k$ is the number of free parameters of the model and 
$n$ is the sample size.

 \begin{figure*}
 \hspace*{1.5076649cm}   
 \vspace*{.1357299cm}   
 \includegraphics[angle=0,scale=0.6]{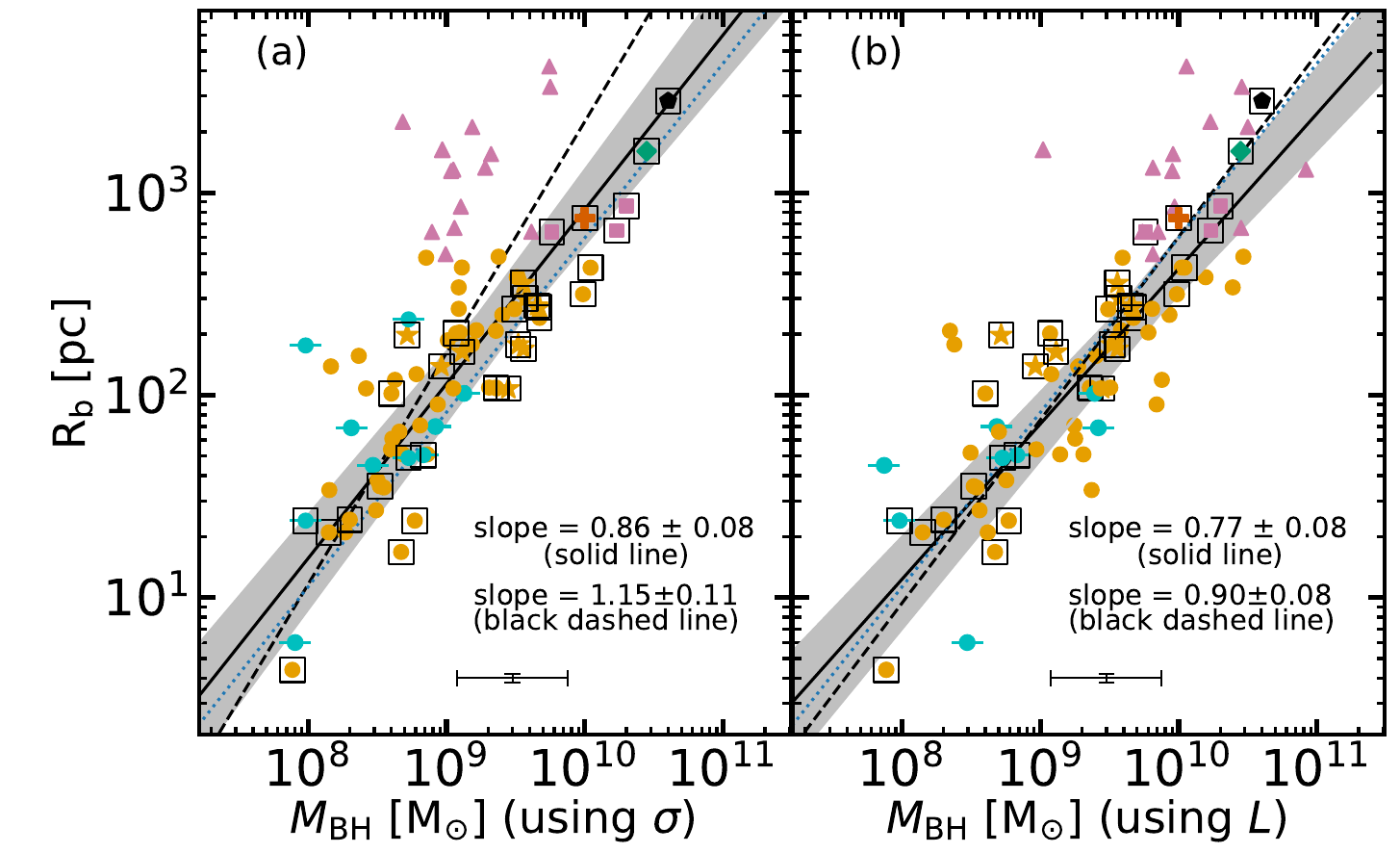}
 \caption{Relationship between $R_{\rm b}$ and $M_{\rm BH}$ for 79
   core-S\'ersic galaxies.  Similar to Fig.~\ref{Fig1}(a), but here,
   besides the 30 core-S\'ersic galaxies with directly measured
   $M_{\rm BH}$, we include 49 (35 normal-core plus 14 large-core)
   core-S\'ersic galaxies with SMBH masses that are predicted using
   the \citet{2013ApJ...764..151G} non-barred $M_{\rm BH}-\sigma$
   relation (a) and their $B$-band core-S\'ersic $M_{\rm BH}-L$
   relation (b).  Symbolic representations are as in
   Fig.~\ref{Fig2}. All core-S\'ersic galaxies with directly measured
   $M_{\rm BH}$ are enclosed in
   boxes. The blue dotted lines represent the
   $R_{\rm b}-M_{\rm BH}$ relation for our full sample of 30
   core-S\'ersic galaxies with measured SMBH masses (see Fig.~\ref{Fig1}a and Table~\ref{Tab2}). The black solid
   lines are the symmetric {\sc ols} bisector regressions for 65
   core-S\'ersic galaxies (excluding the 14 large-core galaxies with
   predicted SMBH masses) and the shaded regions show the associated
   1$\sigma$ uncertainties. The black dashed lines are the symmetric
   {\sc ols} bisector fits to the full sample of 79 core-S\'ersic
   galaxies.  For reference, we also show 4C+37.11 (diamond), Holm 15A (hexagon),
  and NGC 708 (cross).}
 \label{Fig3} 
  \end{figure*}
We use two approaches to derive the AIC and BIC for the single power-law
and broken models: one based on our bisector {\sc bces} fits, and the
other  on the {\sc linmix} fits. For the bisector {\sc bces}-based
models (Table~\ref{Tab4}), we approximate the model evidence using
the residual sum of squares under the assumption of Gaussian
residuals. For the {\sc linmix}-based models (see Appendix), we use the {\sc linmix}
intrinsic scatter (Table~\ref{Tab4}) as part of the Gaussian
likelihood to compute the AIC and BIC. In both cases, we assumed normally
distributed residuals. We find AICc$_{\rm bces,single}$ $-$
AICc$_{\rm bces,broken}$ (= $\Delta$AICc$_{\rm bces}$) $\sim 1.8$ and
AICc$_{\rm linmix,single}$ $-$ AICc$_{\rm linmix,broken}$(=
$\Delta$AICc$_{\rm linmix}$) $\sim 1.6$, which indicate that the
additional complexity of the broken model is not warranted by the
data. Furthermore, the BIC strongly favors the simpler single relation
($\Delta$BIC$_{\rm bces}$ $\sim -4$ and $\Delta$BIC$_{\rm linmix}$
$\sim -7$). Overall, our results suggest that the ($M_{\rm BH}$,
$\sigma$) dataset for the \mbox{non-(large-core)} galaxies is better
described with a single power-law rather than a broken, i.e, two
separate power-laws\footnote{For the {\sc bces}-based single and
  broken models,  $k = 2$ and $k = 4$, respectively. In contrast,
  $k = 3$ and $k = 6$ for the {\sc linmix}-based single and broken
  models, respectively, because the intrinsic scatter is included as
  an additional parameter in the Gaussian likelihood function.}.

Our sample consists of three large-core galaxies with directly
measured SMBH masses (purple boxes), see Figs.~\ref{Fig1}, \ref{Fig2} and
Table~\ref{Tab2}.  We find that these six galaxies are {\it offset upward} by
$\sim (1-2)\times \sigma_{\rm s}$ from the \mbox{$M_{\rm BH}$--$\sigma$}
sequence traced by S\'ersic and normal-core galaxies
(Fig.~\ref{Fig2}). We adopt that 1$\sigma_{\rm s}$ equals the
intrinsic scatter $\epsilon= 0.39$ dex.

In Figs.~\ref{Fig1} and \ref{Fig2}, we also show 4C+37.11(green diamond), 
Holm 15A (black hexagon), and NGC 708 (orange cross), all hosting host UMBHs.  We find that that 4C+37.11, Holm
15A, and NGC 708 lie 1.01, 1.15 and 1.43 dex above the
\mbox{non-(large-core)} \mbox{$M_{\rm BH}$--$\sigma$} relation,
respectively, and are thus offset upward by
$(2.6-3.7)\times \sigma_{\rm s}$ (Fig.~\ref{Fig2}).

While the inclusion of the six large-core galaxies (4C+37.11, NGC~708, NGC~1600, NGC~4486, NGC~4889 and Holm 15A) in the regression analyses
slightly steepens the \mbox{non-(large-core)} $M_{\rm BH}$--$\sigma$
relation (Table~\ref{Tab4}, slope $= 5.28\pm 0.31$), the two slopes
are consistent within 1$\sigma$.

\subsection{The $M_{\rm BH}$--$\sigma$ relation including predicted SMBH masses }\label{Sec4.2b}

In this section, we investigate the $M_{\rm BH}-\sigma$ relation by
including 14 large-core galaxies with BH masses predicted using the
tight \mbox{$M_{\rm BH}-R_{\rm b}$} relation (Fig.~\ref{Fig2}, purple triangles, Section~\ref{Sec2.1}),  and
Table~\ref{Tab4}).  The inclusion of these large-core galaxies
significantly steepens the $M_{\rm BH}-\sigma$ relation
(Fig.~\ref{Fig2}, the solid black line with a slope of
$11.21 \pm 1.58$).  We find that large-core galaxies with
\mbox{$(M_{\rm BH}-R_{\rm b}$)}-based BH masses deviate upward by
$(1-5)\times\sigma_{\rm s}$ from the \mbox{$M_{\rm BH}-\sigma$}
relation traced by \mbox{non-(large-core)} galaxies with directly measured BH masses (Fig.~\ref{Fig2}).
Excluding large-core galaxies, the most deviant outlier in the
$M_{\rm BH}-\sigma$ diagram is the spiral galaxy NGC~5055, see also
\citet{2020ApJ...898...83D}.

\subsection {The $R_{\rm b}$--$M_{\rm BH}$  relation }\label{Sec4.3}

Fig.~\ref{Fig3} plots the core-S\'ersic break radius against black
hole mass for our full sample of 79 core-S\'ersic galaxies. The 30
core-S\'ersic galaxies with directly measured black hole masses are
enclosed in boxes. For the 49  (35 normal-core plus 14 large-core) galaxies with no direct SMBH
measurements, the predicted black hole masses based on $\sigma$ and
$L_{\rm sph}$ are shown in Figs.~\ref{Fig3}a and ~\ref{Fig3}b,
respectively.  Excluding the 14 \mbox{large-core} galaxies with
$\sigma$-based SMBH masses (see Table~\ref{Tab2}), we fit the ordinary
least squares (OLS) bisector regression by \citet{1992ApJ...397...55F}
to the ($R_{\rm b}, M_{\rm BH}$) data of our 65 core-S\'ersic galaxies
(Fig.~\ref{Fig3}a). This fit yields \mbox{log ($R_{\rm b}/$pc) =
  (0.86$\pm$0.08) log ($M_{\rm BH}/10^{9}$)} + (2.06 $\pm$0.05). This
relation (black solid line) is consistent, within the errors, with
that shown in Fig.~\ref{Fig1}(a) based on core-S\'ersic galaxies with
measured $M_{\rm BH}$ (Fig.~\ref{Fig3}, blue dotted lines), although the latter is
vertically shifted by 0.14 dex.

In contrast, Figs.~\ref{Fig3}a shows that fitting the OLS bisector to
the full sample of 79 core-S\'ersic galaxies results in a steeper
slope of $\sim 1.14 \pm 0.11$ and a larger intercept of
$\sim 2.21 \pm 0.10$. The discrepancy is attributed to the
non-large-core (i.e., S\'ersic + normal-core) $M_{\rm BH}-\sigma$
relation, which tend to underestimate SMBH masses in giant spheroids
with large depleted cores (see also \citealt{2007AJ....133.1741B,2007ApJ...662..808L,2011Natur.480..215M,2012ApJ...756..179M,2013ApJ...768...29V,2018MNRAS.474.1342M}).

In Fig.~\ref{Fig3}(b), we find that the non-large-core
$R_{\rm b}-M_{{\rm BH}}$ relation from our OLS bisector fit to 65
core-S\'ersic galaxies (i.e., excluding the 14 \mbox{large-core}
galaxies with $L_{\rm sph}$-based SMBH masses, Table~\ref{Tab2}) has a
slope of $\sim 0.77 \pm 0.07$ and an intercept $\sim 1.86 \pm 0.05$ in
fair agreement with those for the full core-S\'ersic population (slope
$\sim 0.90 \pm 0.08$ and intercept $\sim 1.89 \pm 0.09$).

\section{ Dry mergers and the high-mass upturn}\label{Sec5}

As noted above, the most massive galaxies appear to host BHs that are
overmassive relative to expectations from their central velocity
dispersions, resulting in an upturn in the $M_{\rm BH}$--$\sigma$
relation
\citep[e.g.,][]{2007ApJ...662..808L,2012MNRAS.424..224H,2018MNRAS.474.1342M,2019ApJ...886...80D,2021ApJ...908..134D,2025ApJ...978...48S}. Simulations
by \citet{2006MNRAS.369.1081B} have predicted that merger orbits
become increasingly radial at the highest galaxy masses. In such dry
(nearly parabolic) mergers, orbital energy is converted into internal
energy of the remnant, causing the galaxy to grow in size roughly in
proportion to its stellar mass. Under the virial theorem and energy
conservation, the velocity dispersion is expected to remain largely
unchanged during these events
\citep[e.g.,][]{2003MNRAS.342..501N,2007ApJ...658...65C,2009ApJ...697.1290B,2013ApJ...768...29V,2011MNRAS.412..684B}. Large-core
spheroids are thought to have experienced several ($\sim6$--$10$)
successive major dry mergers \citep{2019ApJ...886...80D}, which add
stars, increase BH mass and enlarge depleted core, while leaving
$\sigma$ approximately constant.

Systematic uncertainties in BH mass measurements may also contribute
to the observed upturn. Dynamically measured BH masses in brightest
cluster galaxies (BCGs) may be biased high owing to limitations in
data quality, assumptions in dynamical modeling, or the treatment of
the dark matter halo
\citep[e.g.,][]{2011Natur.480..215M,2022MNRAS.509.5416T}. In addition,
structural non-homology in the most massive galaxies may introduce
additional scatter or systematic deviations from the standard BH
scaling relations at the high-mass end
\citep[e.g.,][]{2017MNRAS.469.2184F}.

\section{Conclusions }\label{Conc}

Partially depleted cores in luminous, core-S\'ersic galaxies are
thought to form through the transfer of orbital angular momentum from
inspiraling binary SMBHs to stars near the galaxy center, thereby
scouring the core. The core-S\'ersic model describes the underlying
spheroidal light distributions of these galaxies, allowing the size of
the depleted core to be quantified via the break radius ($R_{\rm b}$).
We use a large sample of 151 galaxies with dynamically determined SMBH
masses ($M_{\rm BH}$), to investigate the correlations between
$M_{\rm BH}$ and central velocity dispersion ($\sigma$) and for a
subsample of 30 core-S\'ersic galaxies, between $M_{\rm BH}$ and
$R_{\rm b}$.  We expand the sample of core-S\'ersic galaxies with
directly measured SMBH masses from 24 \citep{2021ApJ...908..134D} to
30, constituting a 25\% increase, and those with predicted SMBH masses
from 27 \citep{2021ApJ...908..134D} to 49, an $\sim$82\% increase. Our
final sample comprises 79 core-S\'ersic galaxies, identified through
detailed modeling of the galaxies' high-resolution {\it HST} surface
brightness profiles.
Our main results are  summarized as follows:\\

1) We find a tight correlation between SMBH mass and core size,
\mbox{$M_{\rm BH} \propto R_{\rm b}^{1.16 \pm 0.10}$}, which holds
across the full mass range of core-S\'ersic, including both
normal-core and large-core spheroids. This updated relation has
Spearman and Pearson correlation coefficients of $ \sim 0.92$, and an
intrinsic scatter of $0.27 \pm 0.07$ dex.  Including three galaxies
recently reported to host UMBHs and large cores, namely 4C+37.11
\citep{2024ApJ...960..110S}, Holm 15A \citep{2019ApJ...887..195M}, and
NGC 708 \citep{2024MNRAS.530.1035D}, does not alter the relation
(\mbox{$M_{\rm BH} \propto R_{\rm b}^{1.15 \pm 0.08}$}), although the
measurement of the break radii for the galaxies is less secure.

2) The $M_{\rm BH}- R_{\rm b}$ relation has a vertical rms scatter in
the \mbox{log $M_{\rm BH}$} of $\Delta_{\rm rms} \sim 0.28$ dex. It is
significantly stronger than the core-S\'ersic $M_{\rm BH}- \sigma$
relation constructed using the same sample, exhibiting $\sim$
30$-$47\% less scatter in the \mbox{log $M_{\rm BH}$}.  The latter has
Spearman and Pearson correlation coefficients of $ \sim 0.79-0.85$ and
a vertical rms scatter of $\Delta_{\rm rms} \sim 0.45 \pm 0.05$ dex.

3) Dividing the sample into S\'ersic, normal-core and large-core
galaxies, we find a single, updated log-linear $M_{\rm BH}-\sigma$
relation for S\'ersic and normal-core galaxies with a slope of
$4.95 \pm 0.29$, \mbox{$\epsilon \sim $0.39 dex} and
\mbox{$\Delta_{\rm rms} \sim 0.46$ dex}.  Including large-core galaxies  with
directly measured BH masses tend to steepen the relation slightly to $5.28 \pm 0.31$, albeit the
slopes are consistent within the errors. While core-S\'ersic galaxies
alone define a much steeper relation with a slope of
$\sim 9.20 \pm 1.53$, our results suggest that S\'ersic and
normal-core galaxies are better described by a common log-linear
$M_{\rm BH}$--$\sigma$ relation, rather than a broken relation. 
  
4) Supermassive and ultramassive black hole masses in large-core
galaxies, whether directly measured or predicted using the
$M_{\rm BH}-R_{\rm b}$ relation, are systematically underestimated by
the $M_{\rm BH}-\sigma$ relation defined by S\'ersic plus normal-core
galaxies.  Three galaxies recently reported to host ultramassive black
holes---4C+37.11(\citealt{2024ApJ...960..110S}), Holm 15A
(\citealt{2019ApJ...887..195M}), and NGC 708
(\citealt{2024MNRAS.530.1035D})---are offset by 1.01, 1.15 and 1.43
dex above the \mbox{non-(large-core)} \mbox{$M_{\rm BH}-\sigma$}
relation, respectively.  Overall, we find that large-core galaxies with
directly measured BH masses are offset upward typically by
$(1-4)\times\sigma_{\rm s}$ from the \mbox{$M_{\rm BH}-\sigma$}
sequence traced by S\'ersic+normal-core galaxies, where
$\sigma_{\rm s}$ denotes the intrinsic scatter (0.39 dex). This
behavior is consistent with a galaxy formation history dominated by
late-time dry mergers, which add stars, increase BH mass, while
leaving $\sigma$ largely unchanged. Our results are consistent with
studies of giant ellipticals and BCGs showing saturation of $\sigma$
at highest luminosities. As such, the high-mass upturn of large-core
galaxies in the $M_{\rm BH}-\sigma$ diagram and the flattening of the
$\sigma-L_{V}$ relation at $M_{V} \la -23.5$ mag
\citep{2019ApJ...886...80D} are internally consistent.

\begin{acknowledgments}
We thank the referee for a timely and constructive report that improved
this manuscript. B.D. is grateful to Jason Aufdenberg for useful comments. 
\end{acknowledgments}

\facilities{HST (WFPC2, ACS, NICMOS, WFC3)}


This work has made use of {\sc numpy} \citep{2011CSE....13b..22V},
{\sc matplotlib} \citep{Hunter:2007}  and {\sc astropy}, a community-developed
core {\sc python} package for Astronomy \citep{2013A&A...558A..33A,2018AJ....156..123A}, {\sc astroquery}
\citep{2019AJ....157...98G} {\sc cubehelix }
\citep{2011BASI...39..289G},
\citep{2016ppap.book...87K}, {\sc scipy} \citep{2020NatMe..17..261V},
and of {\sc topcat} (i.e.\ `Tool for Operations on Catalogues And
Tables', \citealt{2005ASPC..347...29T}) and ChatGPT \citep{OpenAI2023GPT4,OpenAI2025GPT5}.

\appendix 
  
 \setcounter{figure}{0}
\renewcommand{\thefigure}{A\arabic{figure}} 

\setcounter{section}{0}
\renewcommand{\thesection}{A\arabic{section}}

\setcounter{table}{0}
\renewcommand{\thetable}{A\arabic{table}}

\section{{\sc bces} and {\sc linmix} Y$|$X regressions}\label{App2}
Table~\ref{TabA1} presents the BH scaling relations derived from our
forward (Y$|$X) regression fits using {\sc bces}, whereas
Table~\ref{TabA2} presents those obtained from  (Y$|$X) regressions
using {\sc linmix}.
 
\begin{deluxetable*}{ccc}
  \tabletypesize{\scriptsize} \tablewidth{180pt} \tablecaption{Linear Regression Analysis with {\sc bces}   \label{TabA1}} 
  \tablehead{ \colhead{Relation}
    &\colhead{{\sc bces} (Y$|$X) fit} &\colhead{Sample}}
  \colnumbers \startdata
&{\bf Dynamically measured SMBH masses only}&\\
  &{\bf Core-S\'ersic galaxies }&\\
 $M_{\rm BH}-R_{\rm b}$
 &$\mbox{log}\left(\frac{M_{\rm BH}}{M_{\sun}}\right)= (1.08 \pm
           0.10) \mbox{log}\left(\frac{R_{\rm b}}{\mbox{250 pc}}\right)$ +~($9.53 ~ \pm  0.06$)&30 [a]\\
 $M_{\rm BH}-\sigma $
        &$\mbox{log}\left(\frac{M_{\rm BH}}{M_{\sun}}\right)= (9.79  \pm
            17.21) \mbox{log}\left(\frac{\sigma}{\mbox{300 {\rm km s$^{-1}$}}}\right)$ +~($9.53  ~ \pm  0.11$)&30 [a]\\
   &{\bf Including 4C+37.11, Holm 15A, and NGC 708 }&\\
 $M_{\rm BH}-R_{\rm b}$
 &$\mbox{log}\left(\frac{M_{\rm BH}}{M_{\sun}}\right)= (1.06\pm
           0.08) \mbox{log}\left(\frac{R_{\rm b}}{\mbox{250 pc}}\right)$ +~($9.51 ~ \pm  0.05$)&33 [b]\\
$M_{\rm BH}-\sigma $
        &$\mbox{log}\left(\frac{M_{\rm BH}}{M_{\sun}}\right)= (10.40 \pm
            13.47) \mbox{log}\left(\frac{\sigma}{\mbox{300 {\rm km s$^{-1}$}}}\right)$ +~($9.61 ~ \pm  0.16$)& 33 [b]\\
 &{\bf S\'ersic plus Core-S\'ersic galaxies }&\\
 $M_{\rm BH}-\sigma $
 &$\mbox{log}\left(\frac{M_{\rm BH}}{M_{\sun}}\right)= (4.57 \pm 0.31)
 \mbox{log}\left(\frac{\sigma}{\mbox{200 {\rm km s$^{-1}$}}}\right)$
 +~($8.35 ~\pm 0.04$)& 148 [c] \\
$M_{\rm BH}-\sigma $
 &$\mbox{log}\left(\frac{M_{\rm BH}}{M_{\sun}}\right)= (4.69 \pm 0.32)
 \mbox{log}\left(\frac{\sigma}{\mbox{200 {\rm km
         s$^{-1}$}}}\right)$+~($8.37 ~\pm 0.04$)& 151 [d]\\
  $M_{\rm BH}-\sigma $
 &$\mbox{log}\left(\frac{M_{\rm BH}}{M_{\sun}}\right)= (4.83 \pm 0.33)
 \mbox{log}\left(\frac{\sigma}{\mbox{200 {\rm km
         s$^{-1}$}}}\right)$+~($8.39 ~\pm 0.04$)&154 [e]\\\\   \\         
&{\bf Including predicted SMBH masses}&\\
  &{\bf Core-S\'ersic galaxies }&\\
 $M_{\rm BH}-\sigma $
        &$\mbox{log}\left(\frac{M_{\rm BH}}{M_{\sun}}\right)= (9.70 \pm
             1.77) \mbox{log}\left(\frac{\sigma}{\mbox{300 {\rm km s$^{-1}$}}}\right)$ +~($9.82~ \pm  0.10$)& 44 [e]\\
   &{\bf Including 4C+37.11, Holm 15A, and NGC 708 }&\\
  $M_{\rm BH}-\sigma $
        &$\mbox{log}\left(\frac{M_{\rm BH}}{M_{\sun}}\right)= (9.34 \pm
             1.68) \mbox{log}\left(\frac{\sigma}{\mbox{300 {\rm km s$^{-1}$}}}\right)$ +~($9.85~ \pm  0.09$)&47 [f]\\ 
  &{\bf S\'ersic plus Core-S\'ersic galaxies }&\\
  $M_{\rm BH}-\sigma $
 &$\mbox{log}\left(\frac{M_{\rm BH}}{M_{\sun}}\right)= (5.26 \pm 0.36)
 \mbox{log}\left(\frac{\sigma}{\mbox{200 {\rm km s$^{-1}$}}}\right)$
 +~($8.46 ~\pm 0.05$)& 165 [g]\\
     $M_{\rm BH}-\sigma $
 &$\mbox{log}\left(\frac{M_{\rm BH}}{M_{\sun}}\right)= (5.36 \pm 0.36)
 \mbox{log}\left(\frac{\sigma}{\mbox{200 {\rm km s$^{-1}$}}}\right)$
 +~($8.48 ~\pm 0.05$)&168 [i]\\
 \enddata \tablecomments{Similar to Table~\ref{Tab4}, but shown here are the forward  {\sc bces} (Y$|$X) regressions. }
\end{deluxetable*}

\begin{deluxetable}{ccc}
  \tabletypesize{\scriptsize} \tablewidth{80pt} \tablecaption{Linear Regression Analysis with {\sc linmix} \label{TabA2}} 
  \tablehead{ \colhead{Relation}
    &\colhead{{\sc linmix} (Y$|$X) fit} &\colhead{Sample}}
  \colnumbers \startdata
&{\bf Dynamically measured SMBH masses only}&\\
  &{\bf Core-S\'ersic galaxies }&\\
 $M_{\rm BH}-R_{\rm b}$
 &$\mbox{log}\left(\frac{M_{\rm BH}}{M_{\sun}}\right)= (1.04 \pm
           0.11) \mbox{log}\left(\frac{R_{\rm b}}{\mbox{250 pc}}\right)$ +~($9.53 ~ \pm  0.06$)&30 [a]\\
           $M_{\rm BH}-\sigma $
        &$\mbox{log}\left(\frac{M_{\rm BH}}{M_{\sun}}\right)= (8.23 \pm 1.19) \mbox{log}\left(\frac{\sigma}{\mbox{300 {\rm km s$^{-1}$}}}\right)$ +~($9.50  ~ \pm  0.09$)&30 [a]\\
   &{\bf Including 4C+37.11, Holm 15A, and NGC 708 }&\\
  $M_{\rm BH}-R_{\rm b}$
 &$\mbox{log}\left(\frac{M_{\rm BH}}{M_{\sun}}\right)= (1.01\pm
           0.09) \mbox{log}\left(\frac{R_{\rm b}}{\mbox{250 pc}}\right)$ +~($9.52 ~ \pm  0.06$)&33 [b]\\
$M_{\rm BH}-\sigma $
        &$\mbox{log}\left(\frac{M_{\rm BH}}{M_{\sun}}\right)= (9.15 \pm
            1.41) \mbox{log}\left(\frac{\sigma}{\mbox{300 {\rm km s$^{-1}$}}}\right)$ +~($9.60 ~ \pm  0.10$)& 33 [b]\\
 &{\bf S\'ersic plus Core-S\'ersic galaxies }&\\
 $M_{\rm BH}-\sigma $
 &$\mbox{log}\left(\frac{M_{\rm BH}}{M_{\sun}}\right)= (4.59 \pm 0.25)
 \mbox{log}\left(\frac{\sigma}{\mbox{200 {\rm km s$^{-1}$}}}\right)$
 +~($8.38 ~\pm 0.04$)& 148 [c] \\
 $M_{\rm BH}-\sigma $
 &$\mbox{log}\left(\frac{M_{\rm BH}}{M_{\sun}}\right)= (4.70 \pm 0.25)
 \mbox{log}\left(\frac{\sigma}{\mbox{200 {\rm km
         s$^{-1}$}}}\right)$+~($8.39 ~\pm 0.04$)& 151 [d]\\
           $M_{\rm BH}-\sigma $
 &$\mbox{log}\left(\frac{M_{\rm BH}}{M_{\sun}}\right)= (4.84 \pm 0.26)
 \mbox{log}\left(\frac{\sigma}{\mbox{200 {\rm km
         s$^{-1}$}}}\right)$+~($8.41 ~\pm 0.04$)&154 [e]\\\\   \\      
&{\bf Including predicted SMBH masses}&\\
  &{\bf Core-S\'ersic galaxies }&\\
$M_{\rm BH}-\sigma $
        &$\mbox{log}\left(\frac{M_{\rm BH}}{M_{\sun}}\right)= (9.20  \pm
             1.51) \mbox{log}\left(\frac{\sigma}{\mbox{300 {\rm km s$^{-1}$}}}\right)$ +~($9.71~ \pm  0.10$)& 44 [e]\\
   &{\bf Including 4C+37.11, Holm 15A, and NGC 708 }&\\
  $M_{\rm BH}-\sigma $
        &$\mbox{log}\left(\frac{M_{\rm BH}}{M_{\sun}}\right)= (9.10 \pm
             1.55) \mbox{log}\left(\frac{\sigma}{\mbox{300 {\rm km s$^{-1}$}}}\right)$ +~($9.77~ \pm  0.10$)&47 [f]\\ 
  &{\bf S\'ersic plus Core-S\'ersic galaxies }&\\
$M_{\rm BH}-\sigma $
 &$\mbox{log}\left(\frac{M_{\rm BH}}{M_{\sun}}\right)= (5.03 \pm 0.25)
 \mbox{log}\left(\frac{\sigma}{\mbox{200 {\rm km s$^{-1}$}}}\right)$
 +~($8.45 ~\pm 0.04$)& 165 [g]\\ 
      $M_{\rm BH}-\sigma $
 &$\mbox{log}\left(\frac{M_{\rm BH}}{M_{\sun}}\right)= (5.16 \pm 0.26)
 \mbox{log}\left(\frac{\sigma}{\mbox{200 {\rm km s$^{-1}$}}}\right)$
 +~($8.47 ~\pm 0.04$)&168 [i]\\
 \enddata \tablecomments{Similar to Table~\ref{Tab4}, but shown here are the forward  {\sc linmix} (Y$|$X) regressions. }
\end{deluxetable}

\bibliographystyle{aasjournal} 

\bibliography{Bil_Paps_biblo}{}

\end{document}